# A new formulation of protein evolutionary models that account for structural constraints

Andrew J. Bordner[1]* and Hans D. Mittelmann[2]

[1]Mayo Clinic, 13400 East Shea Boulevard, Scottsdale, AZ 85259, USA
[2]School of Mathematical and Statistical Sciences, Arizona State University, P.O. Box 871804, Tempe, AZ 85287, USA

* Correspondence to bordner.andrew@mayo.edu

## Abstract

Despite the importance of a thermodynamically stable structure with a conserved fold for protein function, almost all evolutionary models neglect site-site correlations that arise from physical interactions between neighboring amino acid sites. This is mainly due to the difficulty in formulating a computationally tractable model since rate matrices can no longer be used. Here we introduce a general framework, based on factor graphs, for constructing probabilistic models of protein evolution with site interdependence. Conveniently, efficient approximate inference algorithms, like Belief Propagation, can be used to calculate likelihoods for these models. We fit an amino acid substitution model of this type that accounts for both solvent accessibility and site-site correlations. Comparisons of the new model with rate matrix models and alternative structure-dependent models demonstrate that it better fits the sequence data. We also examine evolution within a family of homohexameric enzymes and find that site-site correlations between most contacting subunits contribute to a higher likelihood. In addition, we show that the new substitution model has a similar mathematical form to the one introduced in (Rodrigue et al. 2005), although with different parameter interpretations and values. We also perform a statistical analysis of the effects of amino acids at neighboring sites on substitution probabilities and find a significant perturbation of most probabilities, further supporting the significant role of site-site interactions in protein evolution and motivating the development of new evolutionary models like the one described here. Finally, we discuss possible extensions and applications of the new substitution model.

## Introduction

Almost all existing amino acid substitution models assume that sites evolve independently. Although this assumption greatly simplifies phylogenetic likelihood calculations, evolutionary selection for a protein to maintain a thermodynamically stable protein structure and conserved fold compatible with its function is expected to induce correlated substitutions between neighboring sites in the three-dimensional structure during the evolutionary history of the protein family. An evolutionary model that accounts for these correlations is expected to be a better null model for detecting evolutionary selection at particular sites. We describe a general formulation of such models based on factor graph representations of probabilistic graphical models, which encode the conditional independence relations between sites induced by local physical interactions between amino acids. We parameterize a particular type of factor graph model for protein evolution, which we refer to as the FG model. Importantly, the use of factor graphs readily allows fast calculations of phylogenetic likelihoods using approximate inference algorithms, such as Belief Propagation (Yedidia et al. 2003), to sum over unobserved intermediate sequences.

We begin by reviewing site-independent protein evolutionary models and compare the amino acid substitution model described here with previous work by (Rodrigue et al. 2005). Next, we review reported experimental and computational evidence for the effects of amino acid interactions on protein evolution and introduce factor graphs and approximate inference algorithms. The general factor graph formulation of protein evolutionary models as well as the particular FG model used in this study are explained in detail in the New Approaches section. As further motivation for the FG model, we present results of a statistical analysis demonstrating that amino acids at nearby sites in the protein structure significantly affect substitution probabilities. We then show results for a comparison of the FG model with alternative models and explain the relation between the present model and the one by (Rodrigue et al. 2005). Finally, we discuss possible extensions and applications of the new substitution model.

## Independent site models of protein evolution

Probabilistic models of protein, as well as DNA, sequence evolution are generally described in terms of a reversible Markov process on a phylogenetic tree. The likelihood of a set of observed protein sequences given all model parameters, including tree topology and branch lengths, can be used to compare alternative models. Significantly, site-independence is almost always assumed in protein evolutionary models. In other words, the substitution probabilities at one site are unaffected by the substitution history at any other site. This greatly simplifies likelihood calculations required for phylogenetic inference and testing since the substitution probabilities can be factorized over sites and calculated in terms of an instantaneous substitution rate matrix. There are several widely used empirical amino acid rate matrices including PAM (Dayhoff et al. 1978), JTT (Jones et al. 1992), WAG (Whelan and Goldman 2001), and LG (Le and Gascuel 2008). In addition, fixed rate models have been extended to account for rate variation across sites by including multiple rate classes, a continuous gamma distribution over rates, and a proportion of invariant sites (Uzzell and Corbin 1971; Tamura and Nei 1993; Yang 1993; Gu et al. 1995). More pertinent to the evolutionary model introduced here, other models that account for protein structure by deriving separate substitution matrices for different local structural environments, defined by *e.g.* discrete solvent accessibility or secondary structure classes, have also been studied and found to improve upon models neglecting structural properties (Goldman et al. 1996; Thorne et al. 1996; Goldman et al. 1998). However, all of these evolutionary models retain the site-independence assumption and therefore ignore potentially significant site-site correlations.

### Relation to previous work

The only previously reported probabilistic models of protein evolution incorporating site-site dependence are the codon substitution model of (Robinson et al. 2003) and the subsequent variant of this model for amino acid substitutions described in (Rodrigue et al. 2005). We refer to the latter as the RO model. The rate matrices in these models were defined through augmenting a traditional independent site substitution rate matrix by multiplying off-diagonal matrix elements by a Boltzmann-type factor, $\exp\left(p\left(E\left(S_i\right)-E\left(S_f\right)\right)\right)$, in which $E(S_i)$ and $E(S_f)$ are the empirical potentials for initial and final amino acid sequence $S_i$ and $S_f$, respectively, and p is a parameter. This functional form incorporates the effect of protein structure and stability by insuring higher rates for substitutions that are expected to improve protein stability, as reflected by a reduced empirical energy $E(S)$. More specifically, the model of (Robinson et al. 2003) included two different empirical potentials, one depending on solvent accessibility and the other depending on site pairs, along with two corresponding parameters. In contrast, the RO model used a single empirical potential depending only on site pairs. Both of these models were formulated within a Bayesian probability framework, in which posterior parameter distributions were estimated based on prior distributions.

As we show below, the FG model has a similar mathematical form to that of the RO model, however with different interpretations of the model parameters and a different normalization scheme. One advantage of the FG model is that it is based on a general probabilistic principle, namely conditional independence of sites that are widely separated in the three-dimensional protein structure given the amino acids at neighboring sites. This enables the model to be formulated using factor graphs, which provide a systematic approximation of this general correlation structure for substitution probabilities. In contrast, the two previous models assumed a particular functional dependence of the rates on the sequence-structure compatibility quantified by the empirical potential, namely the Boltzmann-type factor. Although this dependence is qualitatively plausible, there is no theoretical support for this particular form for the dependence of the substitution rates on the protein structure. More importantly, the FG model allows a more flexible parameterization because all parameters are structure-dependent and fit to homologous sequence data while the other models contain only one or two structure-dependent parameters fit to this type of data. Another advantage of the FG model is that likelihood calculations can be performed using fast approximate inference algorithms, which are introduced below. On the other hand, likelihood calculations in the previous models were performed by summing over substitution histories using a Markov Chain Monte Carlo (MCMC) sampling procedure. One advantage of the previous models is their use of Bayesian prior distributions for the explicit model parameters, while the FG model uses maximum likelihood point estimates of model parameters. Although the FG model could also potentially be formulated within a Bayesian framework, this would require computationally costly summation over prior parameter distributions.

### Biophysical constraints on protein evolution and site-interdependence

Non-synonymous mutations in protein coding regions of DNA directly affect the protein product by changing its amino acid sequence. Observations from a number of studies suggest that protein evolution is profoundly influenced by the biophysical effects of such non-synonymous mutations. These effects can include changes in a protein's biochemical function, stability, and aggregation propensity. Mutations at almost all sites in a protein affect its stability and aggregation (Alber 1989; Matthews 1995; Goldberg 2003). Experimental evidence to date shows that proteins are only marginally stable with thermodynamic stability ($\Delta G$) values in the range between -3 and -10 kcal/mol (Pace 1975; Plaxco et al. 2000), which is comparable to the energy of a single hydrogen bond (Creighton 1992). The former lower limit on stability is easy to understand since low stability leads to a large fraction of unfolded non-functional proteins that are either rapidly degraded or lead to harmful aggregates. The latter upper limit on stability is less well understood and has been explained as a loss of activity due to increased rigidity that prevents functional protein motions (Somero 1995; DePristo et al. 2005) or as a consequence of a mutation/selection/drift steady state based on population genetics models (Wylie and Shakhnovich 2011). Furthermore, differences in $\Delta G$ between the mutant and wild type forms of the protein, or $\Delta\Delta G$, for single residue mutants are mostly within the range of 0.5 – 5 kcal/mol (Alber 1989; Pakula and Sauer 1989; Shortle 1989; Milla et al. 1994; Matthews 1995). The similar magnitudes of $\Delta G$ and $\Delta\Delta G$ suggest that most single residue mutations significantly affect protein stability. This is supported by experimental studies showing that most single residue mutations lead to reduced protein stability (Pakula et al. 1986; Schultz and Richards 1986; Milla et al. 1994). Importantly, recent experimental evidence suggests that misfolded proteins impose an evolutionary fitness cost regardless of the protein's function in the cell (Geiler-Samerotte et al. 2010).

Likewise, sequence analyses also support the significance of physical constraints on protein evolution. First, the predominantly low ratio of non-synonymous to synonymous evolutionary rates, *i.e.* $d_N/d_S < 1$, in protein-coding regions of genes is evidence of purifying selection (Li 1997). Second, there are generally statistically significant numbers of correlated mutations between interacting residues in protein structures (Choi et al. 2005). The observation that these mutations are significant for residues directly interacting through side chain interactions but not those with side chain-backbone interactions, which are mostly independent of the identity of one residue, suggests that local correlations are predominantly due to direct physical interactions (Choi et al. 2005). Also, global analyses of correlated residue substitutions have found that a statistically significant fraction of correlated sites separated in the primary sequence are within interaction distance in the protein structure (Gobel et al. 1994; Shindyalov et al. 1994; Olmea and Valencia 1997; Larson et al. 2000; Singer et al. 2002; Gloor et al. 2005; Vicatos et al. 2005; Kundrotas and Alexov 2006). Connected chains of neighboring sites with correlated residue substitutions are also postulated to mediate the transmission of long-range allosteric effects in proteins (Lockless and Ranganathan 1999; Hatley et al. 2003; Suel et al. 2003; Shulman et al. 2004; Dima and Thirumalai 2006). Finally, both computational (Bornberg-Bauer and Chan 1999; Bloom et al. 2006) and experimental (Bershtein et al. 2006; Bloom et al. 2006) evidence supports the view that increased stability promotes the evolvability of proteins to acquire new functions (Wang et al. 2002; Tokuriki et al. 2008), providing another mechanism by which thermodynamic stability influences the evolutionary histories of protein families.

### Probabilistic graphical models and factor graphs

The new class of protein evolutionary models are formulated in terms of a probabilistic graphical model as represented by a factor graph (Kschischang et al. 2001). A factor graph can be used to represent a joint probability density function that factorizes as

$$p\left(X_1, X_2, \ldots, X_N\right) = \frac{1}{Z} \prod_{j=1}^{m} \phi_j\left(S_j\right),$$

(1)

in which each non-negative factor function $\phi_j$ depends only on a subset $S_j$ of the random variables $X_1, X_2, \ldots, X_N$. The overall multiplicative factor of $1/Z$ insures proper normalization. A factor graph model can be described by a bipartite graph in which random variables and factor functions are represented by two classes of nodes with edges connecting a node for factor function $\phi_j$ with a node for random variable $X_i$ if and only if $\phi_j$ depends on $X_i$. The factorization of the joint PDF in **Equation 1** implies specific conditional independence conditions that can be directly read off from the corresponding factor graph. In the context of our residue substitution model, this conditional independence arises from the assumption that only amino acid substitutions at neighboring sites in the three-dimensional structures are correlated at short evolutionary distances.

### Approximate inference algorithms

Given a factor graph, a common task is to calculate a marginal distribution, which corresponds to a likelihood for the FG evolutionary models described here. Exact calculation of the marginal probabilities by, *e.g.* the Junction Tree algorithm (Lauritzen 1988), is computationally infeasible for the size of factor graph models encountered in this study. Instead, we employed Belief Propagation (Yedidia et al. 2003), a so-called message passing algorithm in which intermediate variables, called messages, are iteratively passed along the graph edges. It generalizes many independently developed special purpose algorithms including the forward-backward algorithm (Rabiner 1990), Kalman filtering (Kalman 1960; Kalman and Bucy 1961), and Felsenstein's pruning algorithm (Felsenstein 1981). The Belief Propagation algorithm is exact for trees but yields an approximate result for other graphs. Although the exact convergence conditions are only known for special classes of graphs (Weiss 2000; Mooij and Kappen 2007), it is guaranteed to converge to at least a local maximum of the posterior probability (Weiss 2000). Belief Propagation has been successfully used for diverse applications including error correction coding for communications (McElice et al. 1998) and image processing (Sun et al. 2003; Felzenszwalb and Huttenlocher 2006). Furthermore, we previously found that the algorithm rapidly converged and yielded accurate results for protein design problems (Bordner 2010). In this work, we also used another approximate inference algorithm, Tree Expectation Propagation (Minka and Qi 2004), for non-trivial phylogenetic trees containing more than two taxa because it converged faster for the resulting larger factor graph inference problems.

## New Approaches

### General formulation of factor graph models of protein evolution

We first describe the general formulation of evolutionary models based on factor graphs and then give details in the next section for the pairwise FG model studied in the remainder of this paper. The basic idea of these models is to express the joint probability of two amino acid sequences at short evolutionary distances, at which the probability of multiple substitutions is small, by a factor graph. This choice is motivated by evidence both from studies described in the Introduction as well as the statistical analysis results described below, which show that substitution probabilities at a particular site are influenced by amino acids at nearby sites in the protein structure. These interactions are presumably mediated by direct physical interactions between amino acids at these proximal sites. This interdependence is encoded in the factor graph by only including factors that depend on amino acids at neighboring sites, *i.e.* excluding factors that depend on sites that are widely separated in the three-dimensional protein structure. A fundamental assumption of these models is that the protein backbone structure, or fold, is conserved for all proteins in the family. This assumption is supported by an analysis of high-resolution protein structures with similar amino acid sequences (Chothia and Lesk 1986; Flores et al. 1993), although proteins in some families adopt multiple structures associated with distinct functional states (Kosloff and Kolodny 2008). The assumption of a conserved fold insures that the set of interacting sites remains constant for a given protein family. The conditional independence properties of factor graphs imply that, at short evolutionary distances, the conditional substitution probability at a site is independent of which amino acids are present at distant sites in the protein structure given the amino acids at neighboring sites. Generally speaking, it enforces locality of the site-site dependencies, resulting in a computationally tractable model. Likelihoods over longer evolutionary distances are calculated by combining multiple copies of the short evolutionary distance factor graph, such that the total evolutionary distance is the sum of distances corresponding to the short-distance factor graphs, to form a larger composite factor graph. Once the complete factor graph for a particular set of sequences and phylogenetic tree is defined, then its likelihood can be calculated using approximate inference algorithms, such as the Belief Propagation algorithm described above, to sum over all possible unobserved intermediate sequences. In principle, Markov Chain Monte Carlo (MCMC) methods can also be employed for likelihood calculations (Winkler 2006), although finding an efficient sampling scheme is expected to be difficult.

## Pairwise factor graph model of protein evolution

We next derive a relatively simple version of the factor graph evolutionary models, called the FG model, which contains only factors depending on one or two random variables. Such an undirected probabilistic graphical model is also referred to as a (pairwise) Markov Random Field.

We begin by defining a model that describes protein sequence evolution over a short evolutionary distance $\Delta t$ and then show that the full FG model over a longer distance has the same functional form. The resulting FG model accounts for site-site correlations as well as the effects of solvent accessibility on amino acid substitution rates. We consider a consensus protein backbone structure with $N$ sites for all sequences in the protein family. Furthermore, let $S^{(m)} \equiv S_1^{(m)} S_2^{(m)} \cdots S_N^{(m)}$ denote an amino acid sequence at distance $t_m$ along a branch consisting of the 20 natural amino acids. For simplicity, we do not explicitly account for deletions or insertions in the model since they would correspond to changes in the protein backbone structure. Instead, we simply treat them as missing data and sum over all possible amino acids at gaps when calculating the likelihood. We also define a set of interacting pairs of sites, $K$, that are nearby in the protein structure. As explained in the Methods section, these are defined by contacting side chains in the representative protein structure taken from the Protein Data Bank (PDB). The short evolutionary distance factor graph model then gives the joint probability of observing amino acid sequences $S^{(m)}$ and $S^{(m+1)}$ in a protein family separated by evolutionary distance $\Delta t$ as

$$p\left(S^{(m)}, S^{(m+1)} \mid \Delta t\right) = \frac{1}{Z_D} \prod_{i=1}^{N} \left(D_1\left(S_i^{(m)}, S_i^{(m+1)}, A_i\right) + \Delta t D_2\left(S_i^{(m)}, S_i^{(m+1)}, A_i\right)\right)$$
$$\times \prod_{(i,j) \in K} C\left(S_i^{(m)}, S_j^{(m)}, A_{ij}\right) C\left(S_i^{(m+1)}, S_j^{(m+1)}, A_{ij}\right)$$
, (2)

subject to the following constraints for all amino acids $s_i$ and $s_j$ and residue property values $a$:

1. $D_1\left(s_i, s_j, a\right) = 0$ if $i \neq j$, (3)

2. $D_2\left(s_i, s_j, a\right) = D_2\left(s_j, s_i, a\right)$,

3. $C\left(s_i, s_j, a\right) = C\left(s_j, s_i, a\right)$

4. $D_1\left(s_i, s_i, a\right) \geq 0$,

5. $D_1\left(s_i, s_j, a\right) + \Delta t_{\max} D_2\left(s_i, s_j, a\right) \geq 0$,

6. $C\left(s_i, s_j, a\right) \geq 0$,

7. $D_1\left(Ala, Ala, a\right) = 1.0$,

8. $C\left(Ala, Ala, a\right) = 1.0$,

in which $\Delta t_{max}$ is the maximum $\Delta t$ value for which the short distance model is defined (0.05), $A_i$ is a structural property at site $i$, and $A_{ij}$ is a structural property shared by sites $i$ and $j$. For the FG model, we chose $A_i$ to be a binary variable that is 1 if the amino acid in the representative structure is buried, as defined by relative SASA < 20%, and 0 otherwise. Likewise, we defined the binary variable $A_{ij} \equiv A_i \wedge A_j$, *i.e.* it represents whether or not amino acids at both sites $i$ and $j$ are buried. The FG model contains three factor functions, $D_1(s_i,s_j,a)$, $D_2(s_i,s_j,a)$, and $C(s_i,s_j,a)$, in which $s_i$ and $s_j$ are the amino acids at the corresponding factor graph nodes and $a$ is the binary solvent accessibility class for the respective site ($D_1$ and $D_2$) or site pair ($C$). The combination of factors $D_1\left(S_i^{(m)}, S_i^{(m+1)}, A_i\right) + \Delta t D_2\left(S_i^{(m)}, S_i^{(m+1)}, A_i\right)$ accounts for a substitution from amino acid $S_i^{(m)}$ to $S_i^{(m+1)}$ at site $i$, which has solvent accessibility $A_i$. The factors $C\left(S_i^{(m)}, S_j^{(m)}, A_{ij}\right)$ and $C\left(S_i^{(m+1)}, S_j^{(m+1)}, A_{ij}\right)$ account for the effects of interactions between amino acids at neighboring sites $i$ and $j$ for the initial and final sequences, respectively. Constraint 1 in **Equation 3** ensures that $\lim_{\Delta t \to 0} p\left(S^{(1)}, S^{(2)}, \Delta t\right) = 0$ if $S^{(2)} \neq S^{(1)}$; Constraint 2 enforces reversibility; Constraint 3 implies that the order of pairs of sites is unimportant; and Constraints 4-6 requires that all factors are positive semi-definite, as is needed to ensure that their normalized product is a well-defined probability density function (see **Equation 1**). Because multiplicative scaling of all $D_1(s_i,s_j,a)$ and $D_2(s_i,s_j,a)$,variables or all $C(s_i,s_j,a)$ variables for a fixed value of $a$ leaves the right-hand side of **Equation 2** unchanged, we introduced Constraints 7 and 8 in order to remove these spurious degrees of freedom and thereby facilitate parameter optimization. The number of independent parameters for each factor can be calculated from the number of different amino acids (20), number of different solvent accessibility classes (2), and accounting for the constraints. The total number of independent parameters for factors $D_1$, $D_2$, and C are 38, 420, and 418, respectively.

Notably, this model is manifestly reversible. Since, according to the first constraint $D_1$ is diagonal we define $d_1(i,a) \equiv D_1\left(i,i,a\right)$. We assume that the factors corresponding to evolution over distance $\Delta t$ at each site have a linear dependence on $\Delta t$, namely they are $D_1 + \Delta t D_2$, as is expected from a short distance approximation in which $O(\Delta t^2)$ terms are negligible. On the other hand, the site pair factors, $C$, were chosen to be independent of evolutionary distance. We also tried fitting linearly dependent factors but found that they were approximately independent of $\Delta t$ (data not shown) and therefore adopted the simpler choice of constant factors.

We next derive the corresponding factor graph model describing the stationary probability $p(S)$ from the short-distance model by calculating the probability for a sequence $S$ to remain unchanged in the limit, $\Delta t \to 0$,

$$\lim_{\Delta t \to 0} p(S, S, \Delta t) = p(S \mid S, \Delta t) p(S) = p(S). \tag{4}$$

Replacing the probability distribution on the left-hand side by the factor graph model expression in **Equation 2** gives

$$p(S) = \frac{1}{Z_S} \prod_{i=1}^{N} d_1(S_i, A_i) \prod_{(i,j) \in K} C^2(S_i, S_j, A_{ij}). \tag{5}$$

The factor graph corresponding to $p(S)$ is shown in **Figure 1(a)**. Using this expression, the conditional probability then becomes

$$p\left(S^{(m+1)} \mid S^{(m)}, \Delta t\right) = \frac{p\left(S^{(m)}, S^{(m+1)}, \Delta t\right)}{p\left(S^{(m)}\right)}$$

$$= \frac{Z_S}{Z_D} \prod_{i=1}^{N} \left( \mathbf{I}_{S_i^{(m)}, S_i^{(m+1)}} + \Delta t \frac{D_2\left(S_i^{(m)}, S_i^{(m+1)}, A_i\right)}{d_1\left(S_i^{(m)}\right)} \right) \prod_{(i,j) \in K} \frac{C\left(S_i^{(m+1)}, S_j^{(m+1)}, A_{ij}\right)}{C\left(S_i^{(m)}, S_j^{(m)}, A_{ij}\right)}, \tag{6}$$

with $\mathbf{I}$ the identity matrix. It is convenient to define a matrix $\mathbf{Q}^{(a)}$ as

$$Q_{i,j}^{(a)} \equiv \frac{D_2(i, j, a)}{d_1(i, a)}. \tag{7}$$

The transition probability over a longer time scale $T \equiv M\Delta t$ is calculated in terms of $M$ tandem copies of the short time factor graph model by summing over unobserved random variables at intermediate distances as

$$p\left(S^{(M)} \mid S^{(0)}, T\right) = \sum_{S_1} \sum_{S_2} \cdots \sum_{S_{M-1}} p\left(S^{(M)} \mid S^{(M-1)}, \Delta t\right) \cdots p\left(S^{(2)} \mid S^{(1)}, \Delta t\right) p\left(S^{(1)} \mid S^{(0)}, \Delta t\right)$$

$$= \frac{1}{Z} \sum_{S_1} \sum_{S_2} \cdots \sum_{S_{M-1}} \prod_{i=1}^{N} \left( \mathbf{I}_{S_i^{(M-1)}, S_i^{(M)}} + \Delta t \mathbf{Q}_{S_i^{(M-1)}, S_i^{(M)}}^{(A_i)} \right) \cdots \left( \mathbf{I}_{S_i^{(1)}, S_i^{(2)}} + \Delta t \mathbf{Q}_{S_i^{(1)}, S_i^{(2)}}^{(A_i)} \right) \left( \mathbf{I}_{S_i^{(0)}, S_i^{(1)}} + \Delta t \mathbf{Q}_{S_i^{(0)}, S_i^{(1)}}^{(A_i)} \right)$$

$$\times \prod_{(i,j) \in K} \frac{C\left(S_i^{(M)}, S_j^{(M)}, A_{ij}\right)}{C\left(S_i^{(0)}, S_j^{(0)}, A_{ij}\right)}$$

$$\approx \frac{1}{Z} \prod_{i=1}^{N} \left[ \exp\left(T\mathbf{Q}^{(A_i)}\right) \right]_{S_i^{(0)}, S_i^{(M)}} \prod_{(i,j) \in K} \frac{C\left(S_i^{(M)}, S_j^{(M)}, A_{ij}\right)}{C\left(S_i^{(0)}, S_j^{(0)}, A_{ij}\right)}, \tag{8}$$

in which $Z$ is the usual normalization factor. The factor graph corresponding to the conditional probability in **Equation 8** is illustrated in **Figure 1(b)**. Note that the ratios of interacting site pair factors, $C$, cancel for intermediate distance points leaving only the ratio of these factors for the final and initial sequences, $S^{(M)}$ and $S^{(0)}$, respectively. The last line follows from approximation of $(\mathbf{I} + \Delta t \mathbf{Q})^M$ by the matrix exponential $\exp(T\mathbf{Q})$. We calculate the matrix exponential in terms of the eigenvalues of $Q$, $\{\lambda_1,\ldots,\lambda_{20}\}$, and a matrix, $A$, whose rows are the corresponding left eigenvectors, as

$$\exp(TQ) = \mathbf{A}^{-1} diag\left(\exp(T\lambda_1),\ldots,\exp(T\lambda_{20})\right)\mathbf{A} \ . \tag{9}$$

**Figure 2** shows a schematic representation of the factor graph corresponding to an FG model calculation for a four taxon phylogenetic tree. There are $N$ factor graph nodes at each tree node, which are joined together by the factor functions shown in that diagram.

## Results and Discussion

### Statistical analysis of residue substitution rates

In the Introduction, we summarized both computational and experimental evidence that the functional requirement of a protein to fold into a thermodynamically stable three-dimensional structure affects the pattern of sequence evolution within the protein family. It is well established that the local structural environment, in particular solvent accessibility, affects both the rate of evolution as well as the observed amino acid frequencies (Overington et al. 1992; Goldman et al. 1998; Bustamante et al. 2000; Conant and Stadler 2009; Franzosa and Xia 2009; Ramsey et al. 2011). Physical interactions between an amino acid and the surrounding amino acids in the protein structure also determine the effect of a substitution on the overall protein stability. We investigated this from a probabilistic viewpoint by performing a statistical analysis of the degree to which substitution probabilities at a site are affected by the identities of amino acids at neighboring sites. For this analysis we used the same set of multiple sequence alignments from diverse Pfam protein families used to parameterize the FG model, which are described in detail in the Methods section. Also, the same definition of neighboring sites, $K$, was employed, namely pairs of sites for which amino acid side chains have atomic contacts in the representative protein structure from the PDB. Furthermore, we performed the analysis on all sequence pairs within a fixed evolutionary distance interval, defined by sequence differences between 5% and 15%. We chose these evolutionary distances, which are larger than those used to parameterize the FG model, for the statistical analysis in order to collect higher amino acid substitution counts while keeping multiple substitutions reasonably unlikely at the upper sequence dissimilarity limit. Also, as in the FG model parameterization procedure described in the Methods section, we limit the number of sequence pairs per Pfam family to 25 in order to reduce sampling bias.

We iterated over all sites in the data set by accumulating the total number of cases in which amino acid $T_{1i}$ was substituted by $T_{1f}$ in the presence of amino acid $T_2$ at a neighboring site. Only cases in which no substitutions occurred at the neighboring site were considered in order to insure that only the influence of that particular amino acid is being accounted for. For each of the 400 possible combinations of $T_{1i}$ and $T_{1f}$ we constructed a 20 × 2 table where the first column contained the number of cases in which $T_{1i}$ was substituted by $T_{1f}$ with the row corresponding to each possible $T_2$ while the second column contained the number of cases in which $T_{1i}$ was substituted by one of the 19 amino acids other than $T_{1f}$ for each possible $T_2$ at the neighboring site. We then tested whether or not the proportion of cases in which $T_{1i}$ was substituted by $T_{1f}$ versus any other amino acid depended on the amino acid at a neighboring site using a two-sided Fisher exact test. Finally, we applied a Benjamini-Hochberg multiple testing correction (Benjamini and Hochberg 1995) to the p-values and calculated the proportion of amino acid substitutions that were affected by the identity of the neighboring amino acid. A false discovery rate (FDR) cutoff of 5% was used. The results showed that substitution frequencies were affected by the identities of amino acids at neighboring sites for a large fraction, 308/400, of possible amino acid substitutions. Separate analyses of buried and surface-exposed sites similarly gave significant fractions of affected substitutions, 232/400 and 205/400 respectively. Lists of substitutions significantly affected by neighboring amino acids are given in the Supplementary Information. These results confirm that neighboring amino acids do indeed affect substitution frequencies and that the definition of neighboring sites used in this analysis, which is the same as that used for the factor graph model, is a reasonable definition for capturing this dependence.

### Comparison with alternative models on short-distance test set data

We first evaluated the FG model by comparing it with other protein evolutionary models using the independent test set of sequence pairs described in the Methods section, which was not used to fit FG model parameters. These sequence pairs were randomly chosen from diverse Pfam protein families and differed at between 1% and 5% of the aligned sites. One alternative model was fit to the same training data as the FG model, except excluding the residue contact factors $C(\cdot, \cdot, \cdot)$. Therefore it could be expressed as two rate matrices, one each for buried and surface sites, that were fit using maximum likelihood estimation. We denote this as the surface accessibility (SA) matrix model. The RO-like (ROL) model, a factor graph model based on an approximate correspondence with the RO model, was also evaluated. This substitution model is described in detail below. For the other alternative models we used the JTT, WAG, and LG rate matrices with a discrete gamma distributed rates model with four rate categories and either no invariant sites (<rate matrix>+$\Gamma$) or a maximum likelihood estimated number of invariant sites (<rate matrix>+F+$\Gamma$). These maximum likelihood calculations were performed using the PhyML 3.0 program (Guindon et al. 2010) with options "-m <rate matrix> –v e –c 4" for the latter model. Likelihood values were evaluated after optimizing the distance between the two sequences in each pair and optimizing the $p$ parameter for the ROL model and associated rate parameters for the rate matrix models. The median log-likelihood ratios and Akaike Information Criterion (AIC) differences relative to the JTT+F+$\Gamma$ model calculated over all sequence pairs are given in **Table 1**. These results show that the FG model, which accounts for both site solvent accessibility and site-site interactions, was the best since it had the lowest $\Delta$AIC of any model tested (p < 2.2 × $10^{-16}$, Wilcoxon signed-rank test with Bonferroni correction). The SA matrix model, which accounts only for solvent accessibility, had the second lowest $\Delta$AIC. The ROL model, which accounts for correlations between neighboring sites but not solvent accessibility, had the highest $\Delta$AIC of the structure-dependent substitution models. Finally, both versions of the JTT, WAG, and LG matrix models, which do not account for structure, had similar ΔAIC values near zero that were all higher than those for the structure-dependent models, indicating that they underperformed those models.

We also examined if the contribution of site interactions to the model performance depends on the total number of sites. This was done by first dividing the test cases into two groups according to whether the number of sites was small (< 150) or large (≥ 150). Next, we calculated the average differences in log-likelihood values between the factor graph model and SA matrix model within each protein family. Finally, we performed a one-sided t-test on these average differences between the small and large site groups. The result showed that the improvement of the factor graph model due to site interactions was larger for the protein families with the large number of sites (p-value < $2.2 \times 10^{-16}$). One explanation of this difference is that larger protein folds have a larger average number of contacts per site (Bastolla and Demetrius 2005) and therefore inter-site correlations have an increased effect on residue substitution probabilities.

## Comparison with alternative models on phylogenetic trees

We compared the evolutionary models on phylogenetic trees with four taxa so that the maximum likelihood tree topology can be readily determined by selecting from the three possible tree topologies. Approximate inference algorithms are fast enough to enable likelihood calculations with FG model likelihood for larger trees; however, exhaustive optimization over tree topologies becomes problematic because their number grows superexponentially with the number of taxa. In order to avoid potential overfitting, we chose sequences from protein families that were not used to fit the FG model parameters. This provides an estimate reflecting the model's expected performance on novel sequence data. We chose the following four enzyme protein families for testing: glucokinase (PF02685), homogentisate 1,2-dioxygenase (PF04209), cytochrome P450 (PF00067), and pancreatic ribonuclease (PF00074). Four gammaproteobacteria sequences were selected as representative glucokinase sequences, while four mammalian sequences were selected for each of the remaining three protein families (shown in **Table 2**). Significantly, the glucokinase, cytochrome P450, and pancreatic ribonuclease enzymes are expected to function as monomers. On the other hand, the mammalian homogentisate 1,2-dioxygenase proteins likely function as homohexameric complexes, as does the human protein (Titus et al. 2000). An X-ray structure of the human complex, which was used as the representative structure for this family, is shown in **Figure 3**. This family was specifically chosen in order to study the contribution of inter-subunit contacts to phylogenetic likelihoods. The FG model was compared with three alternative substitution models by calculating likelihoods and corresponding AIC values for the four sequences in each family using phylogenetic tree topologies and branch lengths optimized using the same model. The alternative models were the JTT+F+Γ rate matrix model and ROL and SA matrix models. The substitution models were compared by calculating ΔAIC relative to the JTT+F+Γ matrix model. As in the above calculations, the *p* parameter in the ROL model and rate parameters in the JTT+F+Γ model were also optimized. The results, shown in **Table 3**, suggest that the

SA matrix and ROL models perform better than the JTT model (which by definition has ΔAIC = 0) and the FG graph model performs the best out of all four models. The smaller improvements for the FG graph model and other two structure-dependent models compared with the JTT model for the pancreatic ribonuclease sequences is likely a manifestation of the smaller likelihood increases observed for the structure-dependent models over the JTT model when applied to small protein structures (this family has only 113 sites in the representative structure), as was found in the sequence pair analysis described in the last section.

The contributions of subunit contacts to the evolutionary likelihood for the homogentisate 1,2-dixoygenase enzymes were assessed by comparing the likelihoods calculated using an FG model generated using a protein structure containing two subunits of interest extracted from the hexameric complex with a model generated assuming no interactions between the subunits, corresponding to twice the log-likelihood for a single subunit (monomer). The results of this analysis are shown in **Table 4**. The higher likelihoods for the dimer structures compared with non-interacting monomers for the two large interfaces as well as for one of the two small interfaces in the complex suggest that site-site interactions between subunits contribute to the molecular evolution of these multimeric mammalian proteins.

### Relation to the RO model

We next demonstrate that the FG model of protein evolution has a similar mathematical form as the RO model (Rodrigue et al. 2005) although the parameters have different interpretations and therefore different values. The RO model substitution rate matrix, given in Equation 7 of their paper, is

$$R_{S_1,S_2} = \begin{cases} 0 & \text{if } S^{(1)} \text{ and } S^{(2)} \text{ differ at more than one position} \\ Q_{lm} \exp\left(p\left(E\left(S^{(1)}\right) - E\left(S^{(2)}\right)\right)\right) & \text{if } S^{(1)} \text{ and } S^{(2)} \text{ differ only at site i, } S_i^{(1)} = l \text{ and } S_i^{(2)} = m \\ -\sum_{S^{(1)} \neq S^{(2)}} R_{S^{(1)},S^{(2)}} & \text{if } S^{(1)} \text{ and } S^{(2)} \text{ are identical} \end{cases}$$

$$(10)$$

$Q_{lm}$ are elements of a standard independent-site amino acid substitution matrix, which they chose to be the JTT matrix (Jones et al. 1992). Furthermore, E(S) is an empirical potential, for which the one described in (Bastolla et al. 2001) was used, and p is a free parameter. We begin by noting that the requirement that the rate matrix vanishes for amino acid sequences that differ at more than one site simultaneously is the usual assumption of a Poisson process for residue substitutions. The FG model in this paper was explicitly defined to satisfy this constraint by assuming that the short-time factors are linear in the evolutionary distance Δt. Furthermore, the probability of n>1 substitutions occurring in the FG model within evolutionary distance Δt goes as $O(\Delta t^n)$ and thus is arbitrarily small relative to the probability of a single substitution in the limit of Δt → 0. Finally, the diagonal elements of the RO model rate matrix are uniquely defined by the constraint that it conserve probability, or $\sum_{S^{(2)}} R_{S^{(1)},S^{(2)}} = 0$.

Next, we examine the off-diagonal rate matrix elements corresponding to sequences differing by one residue substitution. Although the potential of (Bastolla et al. 2001) contains only site pair terms, we consider the more general case of empirical potentials that also include single site terms, or

$$E(S) = \sum_{(i,j) \in K} \varepsilon_{S_i,S_j} + \sum_{i=1}^{N} \varepsilon_{S_i} . \tag{11}$$

Again, $K$ denotes a set of pairs of sites that are nearby in the protein structure, according to some definition that may be different from the one used here. Upon substitution of the empirical potential by **Equation 11**, the rate for a single amino acid substitution at site $i$ in the RO model then becomes

$$\left[ Q_{lm} \exp\left( p\left( \varepsilon_{S_i^{(1)}} - \varepsilon_{S_i^{(2)}} \right) \right) \right] \prod_{(i,j) \in K} \frac{\exp\left( p\varepsilon_{S_i^{(1)},S_j^{(1)}} \right)}{\exp\left( p\varepsilon_{S_i^{(2)},S_j^{(2)}} \right)} . \tag{12}$$

An expression for the corresponding rate in the FG model may be derived by taking the $\Delta t \rightarrow 0$ limit of the conditional substitution probability in **Equation 8** to yield

$$\frac{1}{Z} Q_{lm}^{(A_i)} \prod_{(i,j) \in K} \frac{C\left( S_i^{(2)}, S_j^{(2)}, A_{ij} \right)}{C\left( S_i^{(1)}, S_j^{(1)}, A_{ij} \right)} . \tag{13}$$

The substitution rate for the FG model in **Equation 13** has the same mathematical form as that for the RO model in **Equation 12**. Specifically, they both are the product of a single site term, which for the RO model expression in **Equation 12** is in brackets, and a site-pair term that is itself the product over all interacting sites, $K$, of a ratio of factors for the initial and final amino acid sequences. Taking this connection further, according to the Boltzmann assumption used to derive empirical potentials, the contribution of each pair term to the overall probability that a sequence is compatible with the given structure is $\exp\left( -\varepsilon_{S_i,S_j} / kT \right)$, in which kT is the product of the Boltzmann constant and temperature. On the other hand, according to the expression for this probability in the FG model, shown in **Equation 5**, this contribution is proportional to $C^2\left( S_i, S_j, A_{ij} \right)$. Equating these two expressions for the different models while ignoring overall normalization factors and comparing **Equations 12** and **13** leads to a rough estimate of *1/2* for the $p$ parameter in the RO model, when the empirical potential is expressed in units of kT, as in (Rodrigue et al. 2005). This value is roughly within the range of average posterior values found in that study for different data sets using the JTT rate matrix, $\sim 0.36 < p < \sim 0.63$.

Despite this correspondence between off-diagonal substitution rate matrix elements, the RO and FG models differ in how normalization, which is required for well-defined substitution probabilities, is enforced. As already mentioned, proper normalization of the RO model substitution rates is ensured by defining the diagonal elements of the rate matrix in sequence space as $-\sum_{S^{(1)} \neq S^{(2)}} R_{S^{(1)}, S^{(2)}}$. In contrast, the FG model normalization is imposed by an overall multiplicative factor of 1/Z for the likelihood. It should be emphasized that the normalization terms in both the RO and FG models depend on the protein structure.

We can however use the correspondence between the off-diagonal RO and FG substitution rates to define a so-called RO-like (ROL) model, which is formulated as a factor graph with the same topology as the FG model. According to the discussion above, the ROL model factors are chosen to be

$$
\begin{aligned}
D_1(s_i, s_j, a) &= \delta_{ij} f_i \\
D_2(s_i, s_j, a) &= Q_{ij} f_j \\
S(s_i, s_j, a) &= \exp(-p\varepsilon_{s_i s_j})
\end{aligned}
\qquad (14)
$$

with $\mathbf{Q}$ the JTT substitution rate matrix and $\boldsymbol{\varepsilon}$ the (Bastolla et al. 2001) empirical pair potential parameter matrix. Notably, the factors are the same for a=0 and 1, *i.e.* the model does not account for solvent-accessibility, as is also the case for the RO model. We point out again that this model is not the same as the RO model due to different normalization schemes and the fact that the RO model is formulated in a Bayesian framework while the ROL model calculations are performed using maximum likelihood point estimates of the $p$ parameter and branch lengths.

## Conclusions and Future Work

In this paper, we have introduced a general formulation of amino acid substitution models based on factor graphs. These models can incorporate correlations between neighboring sites and are based on the conditional independence conditions between widely separated sites given the amino acids at neighboring sites, which is expected to be a good approximation due to the significant effects of physically interacting amino acids at nearby sites in the protein structure. A key advantage of these models is that approximate inference algorithms can be used to efficiently sum over evolutionary histories to calculate likelihoods. We have also parameterized and tested a simple version of such models, referred to as the FG model, which contains only pairwise factors. We first derived and parameterized the model for short evolutionary distances and then showed that concatenating these short-distance models to cover longer branch lengths results in a cancellation of the site-site interaction factors, *C*, at intermediate distances, greatly reducing the complexity of the model. We also showed that the FG model is approximately equivalent to the model of (Rodrigue et al. 2005), except for having parameters with different interpretations and consequently distinct values. This correspondence motivated the definition of the ROL model. A comparison of AIC values for sequence pairs and on phylogenetic trees calculated using different amino acid substitution models shows that the FG model fits the sequence data better than the SA matrix model (which only accounts for solvent accessibility), ROL model, or JTT rate matrix model. Finally, we demonstrated that accounting for intersubunit contacts in FG models for a family of multimeric enzymes usually results in higher phylogenetic likelihoods compared with models without these contacts.

There are several possible extensions of the factor graph evolutionary models. One possibility is to construct a model with factors depending on more than two variables, *e.g.* $S_i^{(m)}$, $S_j^{(m)}$, $S_i^{(m+1)}$, $S_j^{(m+1)}$ for $(i,j) \in K$ for sequences $S^{(m)}$ and $S^{(m+1)}$ at distances $t_m$ and $t_{m+1}$, respectively and $\Delta t = t_{m+1} - t_m$ is small. This more detailed model may improve accuracy but would likely require some type of regularization in order to reduce potential overfitting due to the larger number of model parameters. Regularization could be accomplished by adding an $L^1$ penalty term to the likelihood during parameter optimization, as is done in lasso regression (Tibshirani 1996). This would select only a subset of the original model parameters depending on the penalty term weight. The number of parameters could also be reduced by grouping neighboring amino acids into classes defined by their physical properties (*e.g.* nonpolar, uncharged polar, negatively charged, or positively charged). Also, one could formulate alternative models with factors depending on additional structural properties, like secondary structure class. Another possible extension is to derive a corresponding model for RNA sequence evolution, as was done in (Yu and Thorne 2006) based on the DNA codon model of (Robinson et al. 2003). Finally, it would be interesting to see whether factor graph models that account for multiple sets of neighboring sites resulting from different functional protein conformations better fit sequence data for such protein families.

As mentioned in the Introduction, one important application of the FG model is as a null model for detecting negative or positive selection resulting from sources of selection pressure other than maintaining thermodynamic stability of the consensus fold. Furthermore, controlling for phylogeny and structure-dependent effects should improve the detection of non-neutral missense mutations that are associated with disease. Another potential application is the prediction of intersubunit contacts for families of multimeric proteins. This is an outstanding problem because many proteins either have available high-resolution structures or have structures that can be modeled by homology; however, the number of experimental structures of multimeric complexes is limited. Previous methods for predicting site-site contacts are based on multiple sequence alignments and do not make use of phylogenetic trees, *i.e.* they do not account for the shared evolutionary history of extant proteins (Pazos et al. 1997; Weigt et al. 2009; Morcos et al. 2011). In addition, the FG model could potentially be used to detect changes in stoichiometry and thereby aid in studying the evolution of protein-protein interactions. Finally, structure-dependent evolutionary models, like the FG model, are expected to allow more accurate inference of ancestral protein sequences.

## Methods

### Sequence/Structure data set

In order to fit the FG model parameters, we compiled a set of protein structures and aligned sequence pairs from Pfam (Punta et al. 2012) protein families for which at least one high-resolution (≤ 3 Å) structure of a representative family member is available from the Protein Data Bank (PDB) (http://www.pdb.org) (Berman et al. 2000). A single representative PDB structure was selected for each Pfam family based on the criteria of maximum coverage, as defined by the number of aligned amino acids in the structure, and highest resolution. For each Pfam family, we selected a non-redundant set of sequence pairs that differed at between $t_{min}$ = 1% and $t_{max}$ = 5% of the sites using the multiple sequence alignment provided in the Pfam database. Only sequence pairs separated by short evolutionary distances less than $t_{max}$ were selected to fit the short-distance factor graph model because multiple substitutions at a site, which are not accounted for in the linear approximation defining the model in **Equation 2**, are highly unlikely. Excessively short distances less than $t_{min}$ were also avoided because substitution events become too rare to reliably estimate their probabilities. The sequence pairs were chosen so that sequences in different pairs were highly dissimilar, as enforced by a 25% sequence identity cutoff. This reduced redundancy in the training data, which could introduce bias in the model parameter (factor) estimates. Furthermore, in order to more uniformly sample protein families we limited the number of sequence pairs per Pfam family to a maximum of 25, randomly selecting 25 pairs when the number of pairs was larger. Finally, we randomly divided the data into a training set consisting of approximately 90% of the protein families (2469) to be used for fitting the model and a test set containing the remaining 281 protein families. The independent test set was used to evaluate the FG model as well as compare with alternative substitution models.

The set of interacting sites used in the FG model were determined from the representative PDB structures. They were required to have contacting amino acid side chains, as determined by non-hydrogen atom separation ≤ 4 Å, as well as be separated in the primary sequence by at least six residues in order to avoid local contacts due to protein chain connectivity. Because a glycine does not have any non-hydrogen side chain atoms, its $C_\alpha$ atom was considered as a side chain atom for purposes of determining contacts. The set of interacting sites therefore represent sites that are non-local in the protein sequence and where amino acid side chains can physically interact, potentially leading to correlated evolution between those sites. In general, because each amino acid side chain can interact with multiple other side chains a given site can appear in multiple interacting site pairs. The set of interacting site pairs is denoted as $K$ in **Equation 2** above. The solvent accessible surface area (SASA) at each site was calculated with the DSSP program (Kabsch and Sander 1983) using the representative PDB structure for each protein family. The relative SASA was then calculated by dividing by the maximum SASA values taken from (Rost and Sander 1994).

### Parameter estimation using maximum pseudolikelihood

We estimated optimal values for the factors in **Equation 2** by maximizing the pseudolikelihood (Besag 1974) calculated for the training data set, described above, subject to the constraints in **Equation 3**. In general, the pseudolikelihood, $L_{PL}$, for a single data instance with observed values $X_i = x_i$ is

$$L_{PL} = \prod_{i=1}^{N} p\left(x_i \mid N(x_i)\right),\tag{15}$$

in which $N(X_i)$ is the set of random variables which appear in at least one factor function along with $X_i$, *i.e.* their corresponding vertices in the factor graph representation are directly connected by at least one factor node. In terms of the general factor graph definition of **Equation 1** the conditional likelihood in **Equation 15** is

$$p\left(x_i \mid N(x_i)\right) = \frac{\prod_{j:X_i \in S_j} \phi_j\left(x_i, \mathbf{x}_{S_j \setminus X_i}\right)}{\sum_{x_i} \prod_{j:X_i \in S_j} \phi_j\left(x_i', \mathbf{x}_{S_j \setminus X_i}\right)},\tag{16}$$

with $\mathbf{x}_{S_j \setminus X_i}$ denoting values for the other variables besides $X_i$ upon which factor $\phi_j$ depends. Significantly, the pseudolikelihood has the desirable property of statistical consistency, converging to the likelihood as the number of training data increases indefinitely (Gidas 1988; Hyvarinen 2006). Maximizing the pseudolikelihood thus provides a consistent and more computationally practical alternative to maximum likelihood estimation of the factor graph parameters.

### Numerical optimization

The problem of optimizing the pseudolikelihood subject to the variable constraints in **Equation 3** is in the class of non-convex nonlinear optimization. We expressed the problem in the AMPL modeling language (Fourer et al. 2003), which allows a unified interface to different solvers. Global optimization methods are in principle called for but they cannot handle problems of that size. Starting from well-chosen starting values, we applied a local optimization solver. Even that was quite challenging, mostly due to the complexity and nonlinearity of the objective function. We have used KNITRO (Byrd et al. 2006) and from its three algorithms the interior point or barrier method with direct (elimination) linear algebra.

### Phylogenetic likelihood calculations

The FG model likelihood was calculated by generating a file containing the definition of the corresponding factor graph and calculating the marginal probability given the amino acid sequences on the leaf nodes of the phylogenetic tree using a C++ program that calls the libDAI (Mooij 2009) library routines for approximate inference methods. The libDAI library implements both the Belief Propagation and Tree Expectation Propagation algorithms used in this work as well as other approximate inference algorithms that could potentially be used for these calculations.

## Evolutionary model comparison

In order to correct for possible overfitting, the Akaike Information Criterion (AIC) (Akaike 1974) was used for evolutionary model comparisons. The AIC is defined as $AIC = 2k - 2\text{loglik}$, in which *loglik* is the natural log of the maximum likelihood and $k$ is the number of free model parameters that can be varied in the calculation of *loglik*. Because only differences are meaningful, we give the AIC difference ($\Delta$AIC) for each model relative to a reference model, chosen to be JTT+F+$\Gamma$, in **Tables 1** and **3**. Lower $\Delta$AIC values indicate a better model. Because we fit all FG model parameters on an independent training set and only evaluate *loglik* on data from a separate test set, all parameters are fixed in the *loglik* calculation and therefore do not contribute to the AIC penalty term. In general, if one fits a series of increasingly complicated models with increasingly more parameters to a training set then their training set *loglik* values will necessarily be non-decreasing. The AIC penalty term compensates for this, leading to the selection of a model with intermediate complexity. This is likewise the case if one instead evaluates likelihoods on an independent test (holdout) set since they will eventually begin decreasing due to overfitting. The ROL, JTT+$\Gamma$, and JTT+F+$\Gamma$ have 1, 4, and 5 free parameters, respectively, that are optimized on the test set. The AIC values for these models therefore include corresponding penalty terms, which are relatively small. Although the branch lengths are also free parameters, their number does not contribute to $\Delta$AIC because the evolutionary models are compared on phylogenetic trees with the same number of branches.

## Acknowledgements

We thank Joseph Hentz for guidance on statistical tests.

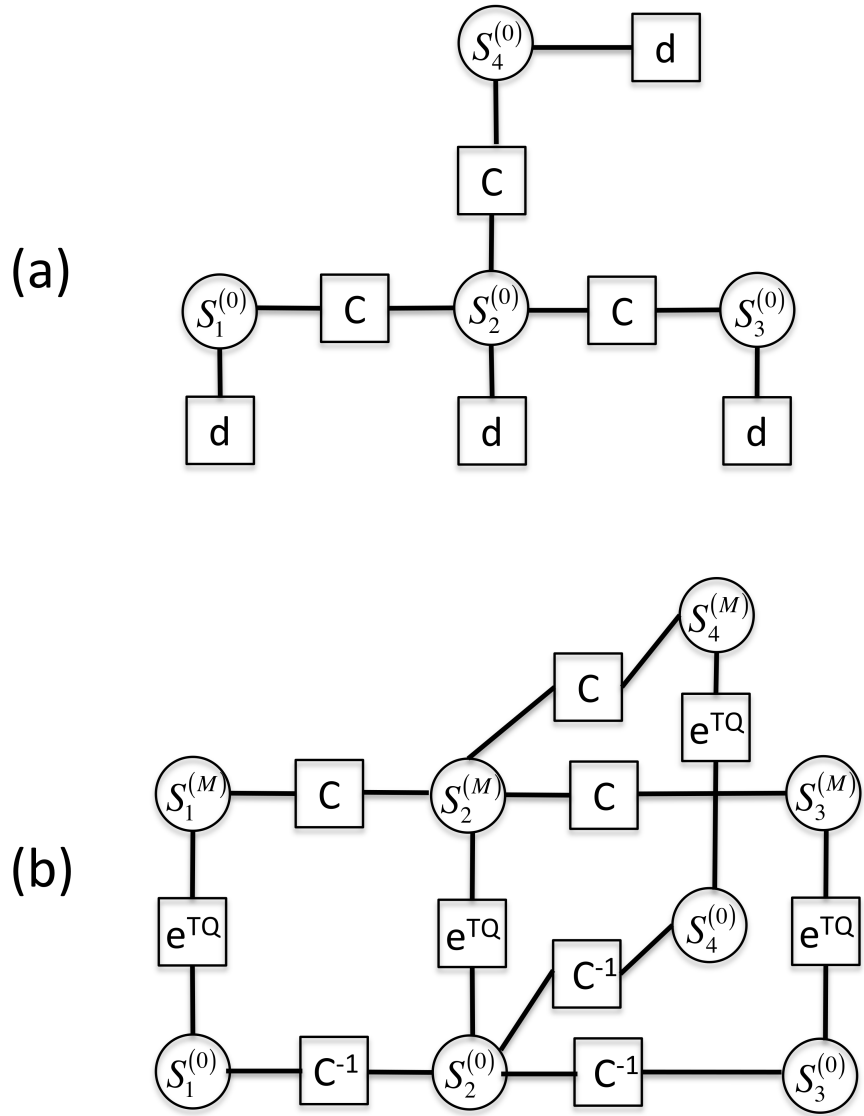

**Figure 1: Factor graphs for a simplified example with only four sites.** Figure 1(a) shows the factor graph for the stationary probability $p\left(S^{(0)}\right)$, given in **Equation 5**, which is evaluated at the root node in likelihood calculations. Figure 1(b) shows the factor graph for the conditional probability $p\left(S^{(M)} \mid S^{(0)}, T\right)$ corresponding to a branch of length T in a phylogenetic tree, as given in **Equation 8**. Nodes, shown as circles, represent the variables, which are the amino acid in the specific sequence at that site. Factors, shown as boxes, are connected to nodes representing dependent variables, which in this case are the amino acids for the particular sequences and sites. The set of neighboring sites for this model is K = {(1,2), (2,3), (2,4)}.

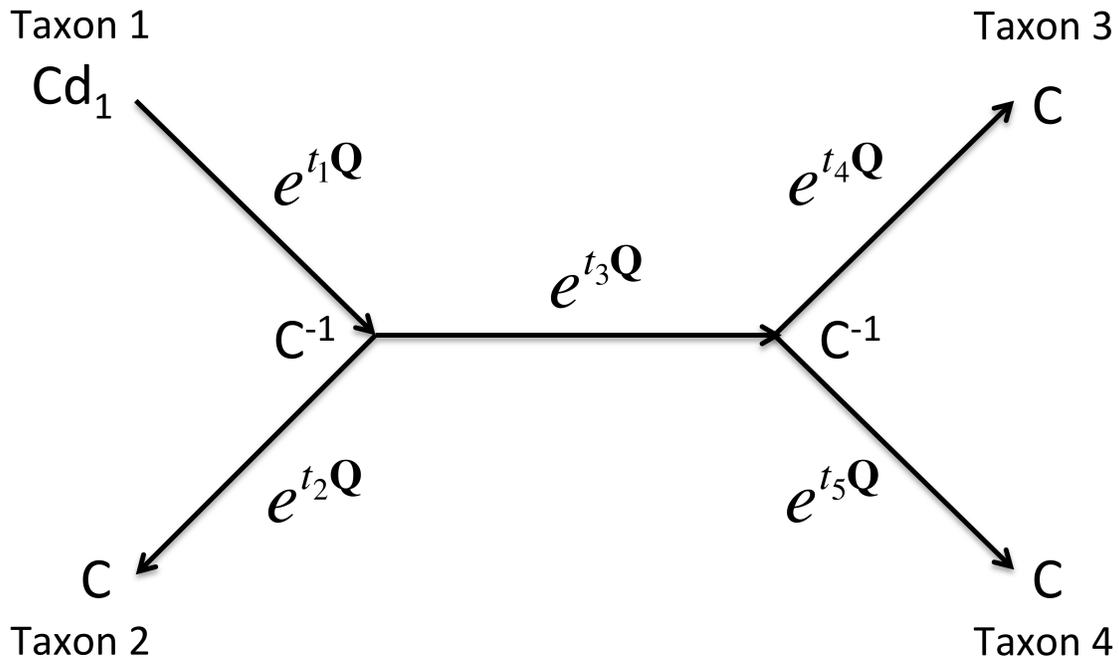

**Figure 2: Schematic representation of the FG model factor graph corresponding to a four taxa phylogenetic tree rooted at taxon 1.** The factor graph has nodes representing the amino acid sequence at each phylogenetic tree node. There is a factor $D(t_j) \equiv \exp\left(t_j \mathbf{Q}^{(A_j)}\right)$ connecting corresponding nodes for each site along a branch having length $t_j$. There are also $C$ factors joining factor graph nodes corresponding to neighboring sites in $K$ at each terminal node and $C^{-1}$ factors at internal nodes. Finally there are $d_1$ factors at each factor graph node corresponding to the tree root node.

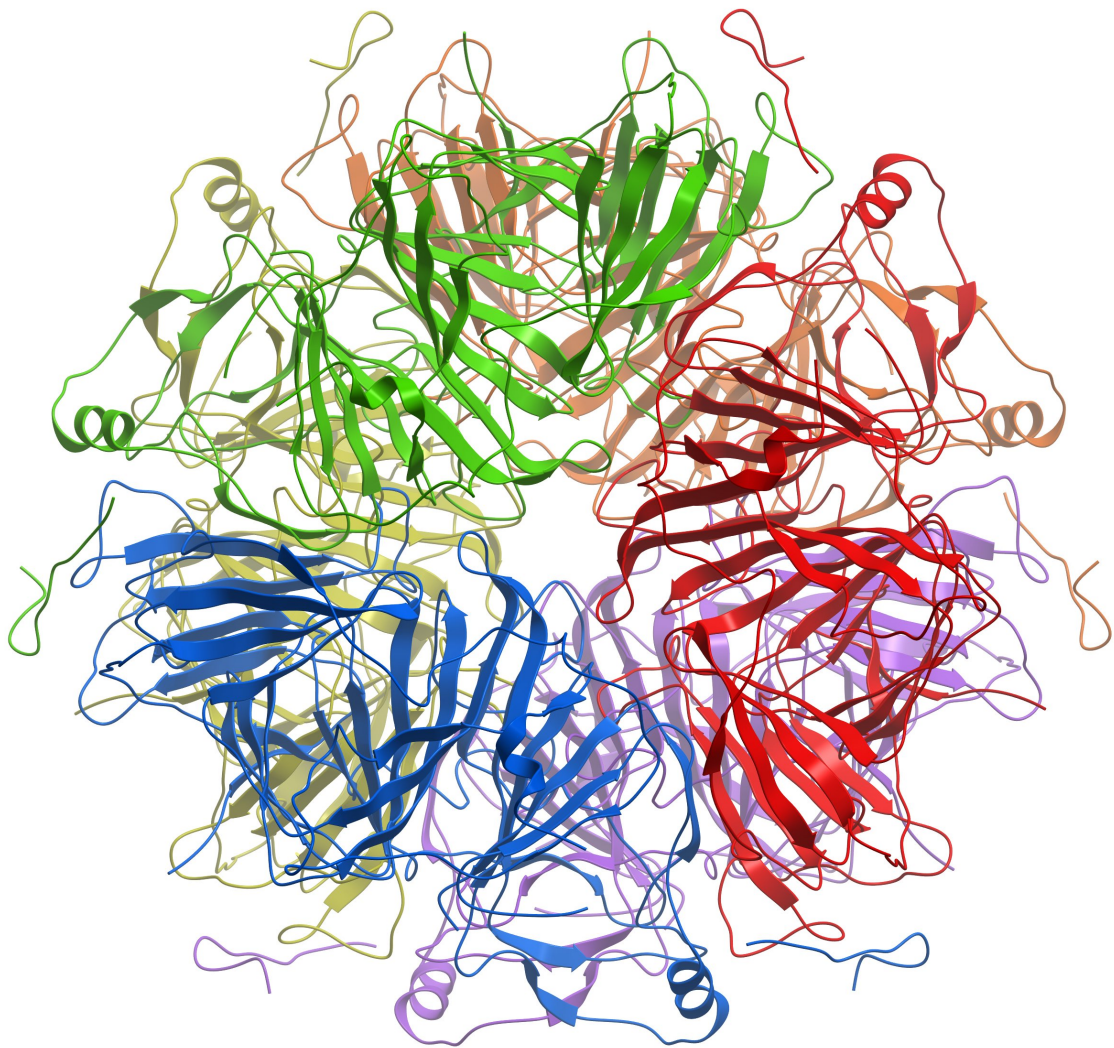

**Figure 3: X-ray structure of human homogentisate 1,2-dioxygenase homohexameric complex (Titus et al. 2000).** Subunits $A_1$-$A_6$, which are defined in **Table 4**, are shown as blue, red, green, yellow, violet, and orange ribbon representations, respectively.

## Tables

**Table 1: Median log-likelihood ratios and AIC differences (ΔAIC) relative to the JTT+F+Γ model for sequence pairs in the test set calculated using different protein evolution models.**

| Substitution model | Median log-likelihood ratio | Median ΔAIC |
|---|---|---|
| FG | 25.8 | -61.6 |
| SA matrix | 11.2 | -32.4 |
| ROL | 9.60 | -27.2 |
| WAG+Γ | -0.775 | -0.450 |
| WAG+F+Γ | -0.774 | 1.548 |
| LG+Γ | -0.360 | -1.28 |
| LG+F+Γ | -0.360 | 0.720 |
| JTT+Γ | $1.50 \times 10^{-4}$ | -2.00 |
| JTT+F+Γ | 0.0 | 0.0 |

**Table 2: Representative structures and sequences for each protein family used in the comparison of evolutionary models.**

| Pfam ID | Protein family | PDB entry (reference) | Sequences Organism (SwissProt entry) |
|---|---|---|---|
| PF02685 | Glucokinase | 1SZ2 (Lunin et al. 2004) | *Yersinia pestis* (GLK_YERPA) *Salmonella enterica* (I0ABJ0_SALET) *Escherichia coli* (GLK_ECO57) *Cronobacter turicensis* (C9XXD4_CROTZ) |
| PF04209 | Homogentisate 1,2-dioxygenase | 1EY2 (Titus et al. 2000) | Human (HGD_HUMAN) Mouse (HGD_MOUSE) Bovine (B8YB76_BOVIN) Rat (Q6AYR0_RAT) |
| PF00067 | Cytochrome P450 | 1Z10 (Yano et al. 2005) | Human (CP2A6_HUMAN) Dog (Q307K8_CANFA) Pig (Q8SQ68_PIG) Mouse (CP2A5_MOUSE) |
| PF00074 | Pancreatic ribonuclease | 1SRN (Martin et al. 1987) | Human (RNAS1_HUMAN) Mouse (RNAS1_MOUSE) Rat (RNAS1_RAT) Horse (RNAS1_HORSE) |

Chain A from the PDB file was used as the protein structure in all cases.

**Table 3: Log-likelihood ratio (LLR) and AIC difference (ΔAIC) for the FG, SA matrix, and ROL models relative to the JTT+F+Γ rate matrix model for the protein families and sequences listed in Table 2.**

| Pfam ID | Protein family | Number of sites | LLR | | | ΔAIC | | |
| | | | SA matrix model | FG model | ROL model (optimal p) | SA matrix model | FG model | ROL model |
|---------|----------------|-----------------|-----------------|----------|-----------------------|-----------------|----------|-----------|
| PF02685 | Glucokinase | 319 | 53.7 | 106.8 | 33.9 (0.81) | -117.4 | -223.6 | -76.8 |
| PF04209 | Homogentisate 1,2-dioxygenase | 635 | 25.8 | 100.0 | 27.3 (0.60) | -61.6 | -210.0 | -62.6 |
| PF00067 | Cytochrome P450 | 320 | 48.2 | 119.6 | 25.6 (0.58) | -106.4 | -249.2 | -59.2 |
| PF00074 | Pancreatic ribonuclease | 113 | 6.68 | 8.10 | 2.56 (0.67) | -23.36 | -26.2 | -13.1 |

**Table 4: Comparisons between the likelihoods calculated using the FG model for the four homogentisate 1,2-dioxygenase taxa with and without specific subunit contacts occurring in the homohexameric human protein complex.**

| Second dimer subunit | Crystallographic symmetry transformation | Interface area ($\text{Å}^2$) | Dimer LLR |
|---|---|---|---|
| $A_2$ | (-y+1,x-y+1,z) | 2375 | 14.0 |
| $A_3$ | (-x+y,-x+1,z) | 2375 | 12.9 |
| $A_4$ | (-y+1,-x+1,-z+½) | 1333 | -0.34 |
| $A_5$ | (x,x-y+1,-z+½) | 1248 | 11.7 |
| $A_6$ | (-x+y,y,-z+½) | 0 | |

All contacts are between the $A_1$ subunit and the specified second subunit. Because the asymmetric unit of the X-ray crystal structure of the protein complex only contains a single subunit, $A_1$, with the other five subunits related by crystallographic symmetry, the corresponding transformations for each subunit are also listed. Subunit $A_6$ does not contact subunit $A_1$ but is included for completeness. The last column shows the log of the ratio between the likelihood calculated using the dimer structure and the likelihood calculated assuming that the subunits are non-interacting monomers, which is simply the square of the monomer likelihood, i.e. dimer LLR = log(likelihood(dimer)) - 2*log(likelihood(monomer)). These results show that three out of the four distinct interfaces in the hexameric enzyme complex are supported by the FG model compared with the null hypothesis of non-interacting subunits, based on the positive log-likelihood ratios.